\documentclass[12pt]{article}
\usepackage{amsthm,amssymb,amsmath}
\usepackage[english]{babel}
 1 
1

\newcommand{\bx}{{\boldsymbol{x}}}

\newcommand{\bR}{{\boldsymbol{R}}}
\renewcommand{\d}{\operatorname{d}}
\newtheorem{teh}{Theorem}

\newcommand{\be}{\begin{equation}}
\newcommand{\ee}{\end{equation}}
\begin{document}

\title{\sc Hydrodynamic reductions and solutions of a
universal hierarchy
}
\author{L. Mart\'{\i}nez Alonso$^{1 }$ and A. B. Shabat $^{2}$ \\
\emph{ $^1$Departamento de F\'{\i}sica Te\'{o}rica II, Universidad
Complutense}\\ \emph{E28040 Madrid, Spain}\\
\emph{$^2$Landau Institute for Theoretical Physics}\\\emph{ RAS,
Moscow 117 334, Russia}}
\date{} \maketitle
\begin{abstract}
The diagonal hydrodynamic reductions  of a hierarchy of integrable
hydrodynamic chains are explicitly characterized. Their compatibility
with previously introduced reductions of differential type is
analyzed and their associated class of hodograph solutions is discussed.
\end{abstract}

\vspace*{.5cm}

\begin{center}\begin{minipage}{12cm}
\emph{Key words:} Hydrodynamic Systems, Differential  Reductions, Hodograph Solutions.

\emph{MSC:} 35L40,58B20.
\end{minipage}
\end{center}
\newpage

\section{Introduction}

In a series of papers \cite{1}-\cite{4} we have considered an infinite
hierarchy of integrable systems \cite{5} which admits many interesting
$(1+1)$-dimensional reductions  like the
Burgers, KdV and NLS hierarchies as well as different classes of
\emph{energy-dependent} hierarchies \cite{6}-\cite{9} including the Camassa-Holm model
also \cite{10} . It motivated the term
\emph{universal hierarchy} we proposed in \cite{1}.

This universal hierarchy can be defined in terms of a generating function $G=G(\lambda,\bx)$
depending on an spectral parameter $\lambda$ and an infinite
set of variables $\bx:=(\ldots,x_{-1},x_0,x_1,\ldots)$, which admits
expansions
\begin{align}\label{1}
G&=1+\frac{g_1(\bx)}{\lambda}+\frac{g_2(\bx)}{\lambda^2}+\cdots,\quad
\lambda\rightarrow\infty,\\
\nonumber \\\label{1p}
G&=b_0(\bx)+b_1(\bx)\lambda+b_2(\bx)\lambda^2+\cdots,\quad
\lambda\rightarrow 0.
\end{align}
The hierarchy is provided by the system of flows
\begin{equation}\label{2}
\partial_n G=\langle A_n,G\rangle, \quad n\in \mathbb{Z},
\end{equation}
where
\[
\langle U,V\rangle:=U(\partial_xV)-(\partial_xU)V,
\]
and
\begin{align}\label{3}
A_n:&=\lambda^n+g_1(\bx)\lambda^{n-1}+\cdots+g_{n-1}(\bx)\lambda+g_n(\bx),\quad n\geq 0\\
\nonumber \\
A_{-n}:&=\frac{b_0(\bx)}{\lambda^n}+\frac{b_1(\bx)}{\lambda^{n-1}}+\cdots
+\frac{b_{n-1}(\bx)}{\lambda},\quad n>0.
\end{align}
In terms of the coefficients $\{g_n(\bx)\}_{n\geq
1}\bigcup\{b_n(\bx)\}_{n\geq 0}$ the system \eqref{2} becomes a
hierarchy of hydrodynamic chains.

An alternative and useful formulation of  \eqref{2} is obtained by
introducing the generating function
\begin{equation}\label{4}
H:=\frac{1}{G}.
\end{equation}
Thus Eq.\eqref{2} is equivalent to
\begin{equation}\label{7}
\partial_n H=\partial_x\Big(A_nH\Big),\quad n\in \mathbb{Z},
\end{equation}
which means, in particular, that the coefficients
$\{h_n(\bx)\}_{n\geq 1}$ supply an infinite set of conservation
laws for \eqref{2}.

The hierarchy \eqref{2} forms a compatible system. Indeed,
as a consequence of \eqref{2} one derives the consistency conditions
\cite{1}-\cite{4}
\begin{equation}\label{8}
\partial_n A_m-\partial_m A_n=\langle A_n,A_m\rangle,\quad n,m \in \mathbb{Z},
\end{equation}

It is important to notice that \eqref{8} implies that the pencil
of  differential forms
\begin{equation}\label{9}
\omega(\lambda):=\sum_{n \in \mathbb{Z}}(H\,A_n)\d x_n,
\end{equation}
is closed with respect to the variables $\bx$ since it satisfies
\begin{align*}
\partial_m(H\,A_n)-&\partial_n(H\, A_m)=\partial_mH\,A_n-
\partial_nH\,A_m+H(\partial_mA_n-\partial_nA_m)\\
=&\partial_x(HA_m)A_n-\partial_x(HA_n)A_m+H\langle A_n,A_m\rangle=0.
\end{align*}
The potential function $Q=Q(\lambda,\bx)$ corresponding to
$\omega$
\begin{equation}\label{10a}
\d Q=\omega,
\end{equation}
leads to another useful formulation of our hierarchy. Indeed, according
to \eqref{9}
\begin{equation}\label{11a}
\partial_n Q=A_n\,\partial_x Q,\quad n\in \mathbb{Z}.
\end{equation}
Moreover, \eqref{11a} is completely determined by $Q$ as
\begin{equation}\label{12a}
G=\frac{1}{\partial_x Q}.
\end{equation}

In Section 2 of this paper the formulation
\eqref{11a}  is applied to prove
that \eqref{2}  includes
multidimensional models such as
\begin{align}
\label{10}
&u_{tx}=u_xu_{yz}-u_{yx}u_z,\\
\label{10p}
&u_{yy}=u_yu_{xz}-u_{xy}u_z,\\
\label{10pp}
&u_{zx}=u_xu_{yy}-u_{xy}u_y,\\
\label{10ppp}
&\Big(\frac{u_t}{u_x}\Big)_t=\Big(\frac{u_y}{u_x}\Big)_z,\\
\label{10pppp}
&\Big(\frac{u_t}{u_x}\Big)_t=\Big(\frac{u_y}{u_x}\Big)_x.
\end{align}

Section 3 deals with the main aim of the present paper:
to characterize the \emph{hydrodynamic
reductions} of \eqref{2}. These reductions are given by the
 solutions of \eqref{2} of the form $G=G(\lambda,\bR)$, where $\bR=(R^1,\ldots,R^{N})$ denotes a finite set of
functions (\emph{Riemann invariants}) satisfying a system of
hydrodynamic equations of diagonal form
\begin{equation}\label{13}
\partial_n R^i=\Lambda ^i_n(\bR) \partial_x R^i,\quad n\in \mathbb{Z}.
\end{equation}
This type of reductions appeared in the context of the dispersionless
KP hierarchy \cite{11}-\cite{14}  and has been also used in
\cite{15} to characterize the integrability of $(2+1)$-dimensional
quasilinear systems. In the present paper we study these
reductions for
the whole set of flows of the hierarchy\eqref{2} and we obtain
their explicit form . Our results are summarized in Theorem 1. It should be
noticed that a similar analysis for
the first member ($t_1$-flow) of \eqref{2} has been recently
performed in \cite{16}.
The  compatibility
between hydrodynamic and differential reductions considered in part 2
of Section 3. The paper finishes with a discussion of the solutions
supplied by the generalized hodograph method \cite{17}.

The following  notation conventions are henceforth used.
Firstly, $G_{\infty}$ and $G_0$ stand for the expansions \eqref{1}
\eqref{1p} of $G$ as $\lambda\rightarrow\infty$ and $\lambda\rightarrow 0$ ,
respectively.  Furthermore,  let $\mathbb{V}$ be the space of
formal Laurent series
\[
V=\sum_{n=-\infty}^{\infty} a_n\lambda^n.
\]
We will denote by $\mathbb{V}_{r,s}$ $(r\leq s)$  the subspaces of elements
\[
V=\sum_{n=r}^{s} a_n\lambda^n.
\]
and by $P_{r,s}:\mathbb{V}\mapsto\mathbb{V}_{r,s}$ the corresponding
projectors. Given $V\in\mathbb{V}$ we will also denote
\[
(V)_{r,s}:=P_{r,s} (V).
\]
In particular, notice that we can write
\[
A_n=\Big(\lambda^n G_{\infty}\Big)_{0,+\infty},\quad
A_{-n}=\Big(\lambda^{-n} G_0\Big)_{-\infty,-1},\quad n\geq 1.
\]

\section{Integrable models arising in the hierarchy}

\subsection{Multidimensional models}

If we use the potential function $Q(\lambda,\bx)$ for the
differential form \eqref{9}
\begin{equation}\label{a1}
\d Q=\sum_{n \in \mathbb{Z}}(H\,A_n)\d x_n,
\end{equation}
then from \eqref{1} one readily deduces that $Q$ admits expansions
of the form
\begin{align}\label{a2}
Q&=\sum_{n\geq 0}\lambda^n
x_n+\frac{q_1(\bx)}{\lambda}+\frac{q_2(\bx)}{\lambda^2}+\cdots,\quad
\lambda\rightarrow\infty,\\
\nonumber \\
Q&=\sum_{n\geq
1}\frac{x_n}{\lambda^n}+p_0(\bx)+p_1(\bx)\lambda+p_2(\bx)\lambda^2+\cdots,\quad
\lambda\rightarrow 0.
\end{align}

By substituting \eqref{a2} into \eqref{a1} and by identifying
coefficients of equal powers of $\lambda$ one obtains formulas for
the differentials of the functions $q_n(\bx)$ and $p_n(\bx)$ in
terms of the coefficients of the expansions \eqref{1} and
\eqref{1p}. For example, the simplest ones are
\begin{align}\label{a3}
\d q_1&=b_0\d x_{-1}+\sum_{n\geq 1}(b_n\,\d x_{-n-1}-g_n\,\d
x_{n-1}),\\ \nonumber \\ \label{a3p}\d p_0&=\frac{1}{b_0}\Big(\d
x_0+\sum_{n\geq 1} (g_n\, \d x_n-b_n\, \d x_{-n})\Big).
\end{align}
They imply
\begin{align}\label{a4}
\d q_1=\frac{1}{\partial_xp_0}\Big(\d x_{-1}-\sum_{n\neq -1}
\partial_{n+1}p_0
\,\d
x_n \Big)\\\nonumber\\
\label{a4p} \d p_0=\frac{1}{\partial_{-1}q_1}\Big(\d x_0-\sum_{n\neq
0}\partial_{n-1}q_1\,\d x_n \Big).
\end{align}
Permutability of crossing derivatives of $q_1$ and $p_0$ in these
identities lead at once to multidimensional nonlinear
equations for the functions $g_n$ and $b_n$. For example,
starting from
\[
\partial_m\partial_{-1}q_1=\partial_{-1}\partial_mq_1,
\quad m\neq -1,
\]
and using \eqref{a4},  the following nonlinear equation results
\begin{equation}\label{a5}
\partial_n\partial_0p_0= \partial_x
p_0\,(\partial_{-1}\partial_{n+1}p_0)-(\partial_{-1}\partial_xp_0)\,D_{n+1}p_0,\quad n\neq -1.
\end{equation}
In the same way, from the crossing relation
\[
\partial_m\partial_nq_1=\partial_n\partial_mq_1,\quad m,n\neq -1,
\]
and \eqref{a4} we get the following nonlinear equation
\begin{equation}\label{a6}
\partial_m\Big(\frac{\partial_{n+1}p_0}{\partial_xp_0}\Big)
=\partial_n\Big(\frac{\partial_{m+1}p_0}{\partial_xp_0}\Big),
\quad m,n\neq -1.
\end{equation}
The same type of equations can be derived for $q_1$. The different
choices available for $n,m$ in the equations \eqref{a5} and
\eqref{a6} give rise to the models \eqref{10}-\eqref{10pppp}.

\subsection{2-dimensional integrable models}

In \cite{1}-\cite{3} we developed a theory of differential
reductions of our hierarchy  based on imposing differential
constraints on $G\approx G_{\infty}$ of the form
\begin{equation}\label{28}
\Big(\mathcal{F}(\lambda,G,G_x,G_{xx},\ldots)\Big)_{-\infty,-1}=0,\quad
x:=x_0.
\end{equation}
In particular the following three classes of reductions associated
to  arbitrary  polynomials $a=a(\lambda)$
 in $\lambda$ were characterized:

\vspace{0.3cm} \noindent {\bf Zero-order reductions}
\vspace{0.3cm}

\begin{equation}\label{r0}
a(\lambda) G=U(\lambda,\bx),\quad U:=\Big(a(\lambda)
G\Big)_{0,+\infty},
\end{equation}

\vspace{0.3cm} \noindent {\bf First-order reductions}
\vspace{0.3cm}

\begin{equation}\label{r1}
G_x+a(\lambda)=U(\lambda,\bx)G,\quad
U:=\Big(\frac{a}{G}\Big)_{0,+\infty},
\end{equation}

\newpage
\vspace{0.3cm} \noindent {\bf Second-order reductions}
\vspace{0.3cm}

\begin{equation}\label{r2}
\frac{1}{2}GG_{xx}-\frac{1}{4}G_x^2+a(\lambda)=U(\lambda,\bx)G^2,\quad
U:=\Big(\frac{a}{G^2}\Big)_{0,+\infty}.
\end{equation}

The first-order reduction for a linear function $a(\lambda)$ determines
the Burgers hierarchy. On the other hand, under the differential constraints \eqref{r2} the hierarchy
\eqref{13} describes the KdV hierarchy and its generalizations
associated to energy-dependent Schr\"{o}dinger spectral problems.
In particular, the linear and quadratic choices for $a(\lambda)$ lead to the KdV
(Korteweg-deVries) and NLS (Nonlinear-Schr\"{o}dinger) hierarchies,
respectively.
Indeed, if we define the functions $\psi(\lambda,\bx)$ by
\begin{equation}\label{s1}
\psi(\lambda,\bx):=\exp(D_x^{-1}\phi),\quad
\phi:=-\frac{1}{2}\frac{H_x}{H}\pm \sqrt{a(\lambda)}\,H,
\end{equation}
then from \eqref{r2} it is straightforward to
deduce that
\begin{align*}
\partial_n\psi&=-\frac{1}{2}(\partial_n\log H)\psi\pm \sqrt{a}A_nH\psi\\&=
A_n(-\frac{1}{2}D_x\log H\pm \sqrt{a}H)\psi-\frac{1}{2} A_{n,x}\psi\\
&=A_n\psi_x-\frac{1}{2}A_{n,x}\psi,\\
\psi_{xx}&=(\phi_x+\phi^2)\psi=(\{D_x,H\}+aH^2)\psi= U\psi.
\end{align*}
In other words, the functions $\psi$ are wave functions for the
integrable hierarchies associated to energy-dependent Schr\"{o}dinger
problems. The evolution law of the potential function $U$ under
the flows \eqref{13} can be determined from the equation
\begin{equation}\label{s2}
\partial_nU=-\frac{1}{2}A_{n,xxx}+2UA_{n,x}+U_xA_n,
\end{equation}
which arises as an straightforward consequence of \eqref{13} and
\eqref{r2}.

Additional reduced hierarchies including nonlinear
integrable models such as  the Camassa-Holm equation can be also
deduced (see \cite{4}).

\section{Hydrodynamic reductions and solutions}
\subsection{Hydrodynamic reductions}

Let us consider now the hydrodynamic reductions of \eqref{2}. We
look for classes a solutions
\begin{equation}\label{3.0}
G=G(\lambda,\bR),
\end{equation}
 of \eqref{2} where $\bR=(R^1,\ldots,R^{N})$
satisfies a infinite system of
hydrodynamic equations of diagonal form \eqref{13}. Our aim is to
characterize both the form of $G=G(\lambda,\bR)$ and the \emph{characteristic
speeds} $\Lambda^i_n$ defining the system \eqref{13}. By substituting
\eqref{3.1} into \eqref{2} then by  using \eqref{13} the identification of
coefficients of the derivatives $\partial_x R^i,\; i=1,\ldots,N$
implies
\begin{equation}\label{3.1}
(D_iG)\Lambda^i_n=A_n(D_iG)-(D_iA_n)G,\quad 1\leq i\leq N,\;\; n\in\mathbb{Z}
\end{equation}
where
\[
\quad D_i:=
\frac{\partial}{\partial R^i}.
\]
In addition to these equations we impose the requirement of
the commutativity of the flows \eqref{13},  which is equivalent to the
following restrictions on the characteristic speeds
\begin{equation}\label{3.2}
\frac{D_j \Lambda_n^i}{\Lambda_n^j-\Lambda_n^i}=
\frac{D_j \Lambda_m^i}{\Lambda_m^j-\Lambda_m^i},\quad
i\neq j,\quad m\neq n.
\end{equation}

  We start our analysis by considering the \emph{positive flows $n\geq 1$}. From
\eqref{3.1} we get the following system for $G\approx G_{\infty}$
\begin{equation}\label{3.3}
(D_i G)\Lambda^i_n=(\lambda^n G)_{-\infty,0}(D_iG)-
(D_i(\lambda^n G)_{-\infty,0})G, \quad n\geq 1.
\end{equation}
By substituting in these equations the expansion $G\approx G_{\infty}$
and identifying the coefficients in $\frac{1}{\lambda}$ and
$\frac{1}{\lambda^2}$ we get
\begin{align}
\label{3.4a}
D_ig_{n+1}&=\Lambda^i_n D_i g_1,\\
\label{3.4b}
D_ig_{n+2}&=\Lambda^i_n D_i g_2+g_{n+1}(D_i g_1)-(D_i g_{n+1})g_1.
\end{align}
As an immediate consequence it follows
\begin{equation}\label{3.5}
\Lambda^i_{n+1}=g_{n+1}+\Lambda^i_n(\Lambda_1^i-g_1),
\end{equation}
which implies
\begin{equation}\label{3.6}
\Lambda^i_n=A_n\Big(\lambda=\Lambda^i_1-g_1\Big),\quad n\geq 1.
\end{equation}

We now look for a system characterizing $\Lambda^i_1$ and $g_1$. To
this end we differentiate \eqref{3.4a} and find
\begin{align*}
&D_jD_ig_{n+1}=D_jD_ig_1\Lambda^i_n+D_i g_1D_j\Lambda^i_n\\
&=D_iD_jg_1\Lambda^j_n+D_j g_1D_i\Lambda^i_n,
\end{align*}
so that
\[
D_{ij}g_1=\frac{D_j\Lambda^i_n}{\Lambda^j_n-\Lambda^i_n}D_ig_1
+\frac{D_i\Lambda^j_n}{\Lambda^i_n-\Lambda^j_n}D_jg_1,
\]
and from \eqref{3.2} we may write
\begin{equation}\label{3.7}
D_{ij}g_1=\frac{D_j\Lambda^i_1}{\Lambda^j_1-\Lambda^i_1}D_ig_1
+\frac{D_i\Lambda^j_1}{\Lambda^i_1-\Lambda^j_1}D_jg_1,\quad i\neq j.
\end{equation}
On the other hand according to \eqref{3.5}
\[
\Lambda^i_2=g_2+\Lambda^i_1(\Lambda^i_1-g_1),
\]
so that from \eqref{3.2} and with the help of \eqref{3.4a} we
find
\[
\frac{D_j\Lambda^i_1}{\Lambda^j_1-\Lambda^i_1}=\frac{D_j\Lambda^i_2}{\Lambda^j_2-\Lambda^i_2}
=\frac{D_j\Lambda^i_1(2\Lambda^i_1-g_1)+(\Lambda^j_1-\Lambda^i_1)D_jg_1}
{(\Lambda^j_1-\Lambda^i_1)(\Lambda^j_1+\Lambda^i_1-g_1)},\quad i\neq j,
\]
which reduces to
\[
D_j\Lambda^i_1=D_j g_1,\quad i\neq j.
\]
In this way
\begin{equation}\label{3.8}
\Lambda^i_1=g_1+f^i(R^i),
\end{equation}
where the functions $f^i$ are arbitrary. By using this result
 in \eqref{3.7}
it follows
\[
D_{ij}g_1=0,\quad i\neq j,
\]
so that
\begin{equation}\label{3.9}
g_1=\sum_{k=1}^N h^k(R^k),
\end{equation}
where the functions $h^k$ are arbitrary.

By using these results we may determine $G(\lambda,\bR)$ since from the
equation \eqref{3.1} with $n=1$
\[
D_i\ln G=\frac{D_i A_1}{A_1-\Lambda^i_1}=\frac{\dot{h}^i(R^i)}{
\lambda-f^i(R^i)},
\]
where $\dot{h}^i:=D_ih^i$. Thus we get
\begin{equation}\label{3.10}
G(\lambda,\bR)=\exp\Big(\sum_{i=1}^N \int^{R^i}\frac{\dot{h}^i(R^i)}{
\lambda-f^i(R^i)}\d R^i\Big),
\end{equation}
where the undefinite integrations are determined up to a  function
of $\lambda$ decaying at $\lambda\rightarrow\infty$. The expression
\eqref{3.10} coincides with the generating function of the conservation
laws densities found in \cite{16}.

 Let us now determine the characteristic speeds $\Lambda^i_{n}$
for the \emph{negative flows $n\leq -1$}. The equations
\eqref{3.1} imply the following system for $G\approx G_{0}$
\begin{equation}\label{3.11}
(D_i G)\Lambda^i_{-n}=(\lambda^{-n} G)_{0,\infty}(D_iG)-
(D_i(\lambda^{-n} G)_{0,\infty})G, \quad n\geq 1.
\end{equation}
Then by inserting the expansion $G\approx G_{0}$ of \eqref{1p}
and identifying the coefficients in $\lambda^0$ and
$\lambda$ we get
\begin{align}
\label{3.12a}
&D_ib_n=(b_n+\Lambda^i_{-n}) D_i\ln b_0,\\
\label{3.12b}
&b_0D_ib_{n+1}=(b_n+\Lambda^i_n) D_i b_1+b_{n+1}(D_i b_0)-
(D_i b_n)b_1.
\end{align}
It implies the following recurrence relation for the characteristic
speeds
\begin{equation}\label{3.13}
\Lambda^i_{-n-1}=\frac{\Lambda^i_{-1}}{b_0}(b_n+\Lambda^i_{-n}),
\end{equation}
which implies
\begin{equation}\label{3.14}
\Lambda^i_{-n}=A_{-n}\Big(\lambda=\frac{b_0}{\Lambda^i_{-1}}\Big),\quad n\geq 1.
\end{equation}
The only unknown now is $\Lambda^i_{-1}$ since according to \eqref{3.10}
\[
b_0=\exp\Big(-\sum_{i=1}^N \int^{R^i}\frac{\dot{h}^i(R^i)}{
f^i(R^i)}\d R^i\Big).
\]
To find $\Lambda^i_{-1}$ we use \eqref{3.13} for $n=1$
\[
\Lambda^i_{-2}=\frac{\Lambda^i_{-1}}{b_0}(b_1+\Lambda^i_{-1}),
\]
and the commutativity condition for the $n=-1$ and $n=-2$ flows
\[
\frac{D_j\Lambda^i_{-1}}{\Lambda^j_{-1}-\Lambda^i_{-1}}
=\frac{D_j\Lambda^i_{-2}}{\Lambda^j_{-2}-\Lambda^i_{-2}}.
\]
Thus by eliminating $\lambda^i_{-2}$ one finds at once
\[
D_j\ln \Lambda^i_{-1}=D_j\ln b_0,\quad i\neq j.
\]
Therefore,
\begin{equation}\label{3.15}
\Lambda^i_{-1}=b_0 \,g^i(R^i),
\end{equation}
with $g^i$ being arbitrary functions.

At this point we have determined all the unknowns of our problem.
However, in our calculation we only used a subset of the equations
required, so we must prove that our
solution satisfies the full system of equations \eqref{3.1}-\eqref{3.2}.

Let us begin with \eqref{3.1} for $n\geq 1$
\[
D_i\ln G=\frac{D_i A_n}{A_n-\Lambda^i_n},\quad n\geq 1,
\]
which according to \eqref{3.6}, \eqref{3.8}-\eqref{3.10} reads
\begin{equation}\label{3.16}
D_iA_n(\lambda)=\frac{\dot{h}^i}{
\lambda-f^i}\Big(A_n(\lambda)-A_n(\lambda=f^i)\Big).
\end{equation}
To prove this identity we express $A_n(\lambda)$ as
\begin{align*}
A_n(\lambda)&=(\lambda^n G)_{0,\infty}=\frac{1}{2\pi i}\int_{\gamma_{\infty}}
\frac{(\widetilde{\lambda}^n G(\widetilde{\lambda}))_{0,\infty}}
{\widetilde{\lambda}-\lambda}\d \widetilde{\lambda}=
\frac{1}{2\pi i}\int_{\gamma_{\infty}}
\frac{\widetilde{\lambda}^n G(\widetilde{\lambda})}
{\widetilde{\lambda}-\lambda}\d \widetilde{\lambda}\\\\
&=\frac{1}{2\pi i}\int_{\gamma_{\infty}}
\frac{\widetilde{\lambda}^n}
{\widetilde{\lambda}-\lambda}
\exp\Big(\sum_{i=1}^N \int^{R^i}\frac{\dot{h}^i}{
\widetilde{\lambda}-f^i}\d R^i\Big)
\d \widetilde{\lambda},
\end{align*}
where $\gamma_{\infty}$ is a positively oriented closed loop around $\infty$ in the
complex plane of $\widetilde{\lambda}$ ($\lambda$ and $f^i$ are assumed
to lie inside the loop). Now by differentiating this
expression with respect to $R^i$ one finds
\begin{align*}
D_iA_n(\lambda)&=
\frac{1}{2\pi i}\int_{\gamma_{\infty}}
\frac{\dot{h}^i\widetilde{\lambda}^n  G(\widetilde{\lambda})}
{(\widetilde{\lambda}-\lambda)(\widetilde{\lambda}-f^i)}
\d \widetilde{\lambda}=\frac{1}{2\pi i}\int_{\gamma_{\infty}}
\frac{\dot{h}^i A_n(\widetilde{\lambda})}
{(\widetilde{\lambda}-\lambda)(\widetilde{\lambda}-f^i)}
\d \widetilde{\lambda}\\\\
&=\frac{\dot{h}^i}{\lambda-f^i}\Big( A_n(\lambda)-A_n(\lambda=f^i)\Big),
\end{align*}
which proves \eqref{3.16}. This identity leads also at once to
\eqref{3.2} since it implies
\[
\frac{D_j A_n(\lambda=f^i)}{A_n(\lambda=f^j)-A_n(\lambda=f^i)}=
\frac{\dot{h}^j}{f^j-f^i},\quad i\neq j,
\]
which means that
\[
\frac{D_j \Lambda^i_n}{\Lambda^j_n-\Lambda^i_n}=
\frac{D_j \Lambda^i_1}{\Lambda^j_1-\Lambda^i_1},\quad n>1.
\]

Let us consider now \eqref{3.1} for $n\leq -1$
\[
D_i\ln G=\frac{D_i A_{-n}}{A_{-n}-\Lambda^i_{-n}},\quad n\geq 1.
\]
From \eqref{3.10}, \eqref{3.14} and \eqref{3.15} it takes the form
\begin{equation}\label{3.17}
D_iA_{-n}(\lambda)=\frac{\dot{h}^i}{
\lambda-f^i}\Big(A_{-n}(\lambda)-A_{-n}(\lambda=\frac{1}{g^i})\Big).
\end{equation}
In order to proof \eqref{3.17} we express $A_{-n}(\lambda)$ in the form
\begin{align*}
A_{-n}(\lambda)&=(\lambda^{-n} G)_{-\infty,-1}=
\frac{1}{2\pi i}\int_{\gamma_0}
\frac{(\widetilde{\lambda}^{-n} G(\widetilde{\lambda}))_{-\infty,-1}}
{\widetilde{\lambda}-\lambda}\d \widetilde{\lambda}=
\frac{1}{2\pi i}\int_{\gamma_0}
\frac{\widetilde{\lambda}^{-n} G(\widetilde{\lambda})}
{\widetilde{\lambda}-\lambda}\d \widetilde{\lambda}\\\\
&=\frac{1}{2\pi i}\int_{\gamma_0}
\frac{\widetilde{\lambda}^{-n}}
{\widetilde{\lambda}-\lambda}
\exp\Big(\sum_{i=1}^N \int^{R^i}\frac{\dot{h}^i}{
\widetilde{\lambda}-f^i}\d R^i\Big)
\d \widetilde{\lambda},
\end{align*}
where $\gamma_0$ is a closed small loop with negative
orientation around  $\widetilde{\lambda}=0$ ($\lambda$ and $f^i$ are
assumed to lie outside the loop ). By differentiating  with
respect to $R^i$ it yields
\begin{align*}
D_iA_{-n}(\lambda)&=
\frac{1}{2\pi i}\int_{\gamma_0}
\frac{\dot{h}^i\widetilde{\lambda}^{-n}  G(\widetilde{\lambda})}
{(\widetilde{\lambda}-\lambda)(\widetilde{\lambda}-f^i)}
\d \widetilde{\lambda}\\\\
&=\frac{1}{2\pi i}\int_{\gamma_0}
\frac{\dot{h}^i A_{-n}(\widetilde{\lambda})}
{(\widetilde{\lambda}-\lambda)(\widetilde{\lambda}-f^i)}
\d \widetilde{\lambda}
=\frac{\dot{h}^i}{\lambda-f^i}\Big( A_{-n}(\lambda)-A_{-n}(\lambda=f^i)\Big),
\end{align*}
Hence \eqref{3.17} holds if we set
\[
g^i(R^i)=\frac{1}{f^i(R^i)}.
\]
The remaining commutativity conditions \eqref{3.2} follow at once.

Therefore we may summarize our analysis in the next theorem

\begin{teh} The hydrodynamic reductions of the hierarchy \eqref{2}
are determined by
\begin{align}\label{3.18}
&G(\lambda,\bR)=\exp\Big(\sum_{i=1}^N \int^{R^i}\frac{\dot{h}^i(R^i)}{
\lambda-f^i(R^i)}\d R^i\Big),
\\\nonumber\\
\label{3.19}
&\partial_n R^i=\Lambda^i_n(\bR)\, \partial_x R^i,\quad \Lambda^i_n(\bR):=A_n(\lambda=f^i(R^i)),
\end{align}
where the functions $h^i$ and $f^i$ are arbitrary.
\end{teh}

Since the systems \eqref{13}  are invariant under local transformations
of the form $R^i\rightarrow\widetilde{R^i}(R^i)$, without loss of generality
we will henceforth set
\begin{equation}\label{gauge}
h^i(R^i)=R^i,
\end{equation}
so that the form of the reduced generating function is
\begin{equation}\label{3.20a}
G(\lambda,\bR)=\exp\Big(\sum_{i=1}^N \int^{R^i}\frac{\d R^i}{
\lambda-f^i(R^i)}\Big),
\end{equation}

In this way some of the simplest hydrodynamic reductions \eqref{3.18} are
given by
\begin{align}
\label{3.20}
&f^i(R^i)=c_i,\quad G=\exp\Big(\sum_{i=1}^N \frac{R^i}{\lambda-c_i}\Big),
\\
\label{3.21}
&f^i(R^i)=-\frac{R^i}{\epsilon_i},\quad G=\prod_{i=1}^N \Big(
\frac{\epsilon_i\lambda+R^i}{\epsilon_i\lambda+\lambda_i}\Big)^{\epsilon_i}.
\end{align}

We also notice that the characteristic velocities of the hydrodynamic systems
 \eqref{3.19} can
be written in terms of the Schur polynomials
\[
\exp(\sum_{n\geq 1}k^n x_n)=\sum_{n\geq 0}k^n S_n(x_1,\ldots,x_n),
\]
as
\begin{align*}
&\Lambda^i_n(\bR)=\sum_{j=0}^n S_j(I_1,\ldots,I_j)\Big(f^i(R^i)\Big)^{n-j},
\quad n\geq 0,\\
&\Lambda^i_{-n}(\bR)=\exp(-I_0)\sum_{j=0}^{n-1} S_j(I_{-1},\ldots,I_{-j})\Big
(f^i(R^i)\Big)^
{-n+j},\quad n>0,
\end{align*}
where
\[
I_n:=sgn(n)\sum_{i=1}^N \int^{R^i}(f^i(R^i))^{n-1}\d R^i.
\]
\subsection{Compatibility with differential reductions}

A natural question is to find the hydrodynamic reductions
compatible with the differential reductions \eqref{r1} and
\eqref{r2}. Let us prove the following result:

\begin{teh}
The only hydrodynamic reductions \eqref{3.20a} compatible
with either first or second order differential constraints are
characterized by
\[
f^i(R^i)=-R^i+c_i,\quad i=1,\ldots,N,
\]
which correspond to generating functions of the form
\[
G(\lambda,\bR)=\alpha(\lambda)\prod_{i=1}^N (\lambda+R^i-c_i).
\]
\end{teh}
\begin{proof}

If we substitute \eqref{3.20a} into the differential constraint \eqref{r2} for second-order
differential reductions we get
\[
\frac{a(\lambda)}{G^2}=U-\frac{1}{2}\sum_i\Big( \dot{f}^i\frac{(\partial_x R^i)^2}
{(\lambda-f^i)^2}+\frac{\partial_{xx} R^i}{\lambda-f^i}\Big)
-\frac{1}{4}\Big(\sum_i\frac{\partial_x R^i}{\lambda-f^i}\Big)^2.
\]
This means that $G$ has a simple zero at each $\lambda=f^i(R^i)$ so that from \eqref{3.20a}
we have
\[
\exp\Big(\int^{R^i}\frac{\d R^i}{\lambda-f^i(R^i)}\Big)=(\lambda-f^i(R^i))H(\lambda,R^i),
\]
where $H$ is different from zero at $\lambda=f^i(R^i)$. If we now differentiate with respect
to $R^i$ we deduce that
\[
\frac{1}{\lambda-f^i}=-\frac{\dot{f}^i}{\lambda-f^i}+\mathcal{O}(1),\quad \lambda\rightarrow\infty.
\]
Therefore the statement of the theorem for second-order differential constraints follows. The
corresponding proof for first-order constraints is similar.
\end{proof}

\subsection{Hodograph solutions}

The general solution of the infinite system \eqref{3.19} is
provided by the implicit \emph{generalized hodograph}
formula \cite{17}
\begin{equation}
\label{3.22}
x+\sum_{n\in \mathbb{Z}-\{0\}}\Lambda^i_n(\bR)\, x_n=\Gamma^i(\bR),
\quad i=1,\ldots,N,
\end{equation}
where the functions $\Gamma^i$ are the general solution of the linear system
\begin{equation}\label{3.23}
\frac{D_j \Gamma^i}{\Gamma^j-\Gamma^i}=
\frac{1}{f^j-f^i},\quad i\neq j.
\end{equation}
By introducing the potential function $\Phi(\bR)$
\[
\Gamma^i=D_i\Phi,\quad i=1,\ldots,N,
\]
the system \eqref{3.23} reduces to the Laplace type form
\begin{equation}\label{3.24}
(f^i-f^j)D_iD_j\Phi=D_i\Phi-D_j\Phi,\quad i\neq j,
\end{equation}
the general solution of which depends on $N$ arbitrary functions
of one variable.

In particular it is immediate to deduce \cite{16} that the generating
function \eqref{3.20a} provides a one-parameter family of
solutions of \eqref{3.24}. Thus we can produce important solutions
of \eqref{3.23} from linear superpositions of $G(\lambda,\bR)$. For
example
\begin{align*}
\Lambda^i_n&=D_i\Phi,\;\; \Phi:=\frac{1}{2\pi i}\int_{\gamma_{\infty}}
\lambda^n G(\lambda,\bR)\d \lambda,\quad n\geq 0,\\
\Lambda^i_{-n}&=D_i\Phi,\;\; \Phi:=\frac{1}{2\pi i}\int_{\gamma_0}
\lambda^{-n} G(\lambda,\bR)\d \lambda,\quad n> 0.
\end{align*}
Furthermore, in several important cases the general solution of
\eqref{3.24} can be written in terms of $G(\lambda,\bR)$.

\vspace{0.2truecm}
\noindent
{\bf Examples }

If we set
\[
f^i(R^i)=\frac{R^i}{n},\quad n=1,2,\ldots,
\]
then the generating function (non normalized as $\lambda\rightarrow\infty$)
is
\[
G(\lambda,\bR)=\prod_{i=1}^N\Big(\lambda-\frac{R^i}{n}\Big)^{-n}.
\]
Hence, by taking for
each $i=1,\ldots,N$  a closed loop $\gamma_i$ in the complex $\lambda$-plane
with positive orientation around $\lambda_i=\frac{R^i}{n}$, the
general solution of \eqref{3.24} can be expressed as
\begin{equation*}
\Phi:=\frac{1}{2\pi i}\sum_{i=1}^N\int_{\gamma_i}
\phi_i(\lambda) G(\lambda,\bR)\d \lambda,
\end{equation*}
where the functions $\phi_i(\lambda)$ are arbitrary.
For example, if $n=1$ it takes the form
\begin{equation*}
\Phi=\sum_{i=1}^N\phi_i(R^i)\prod_{k\neq i}\frac{1}{R^i-R^k}.
\end{equation*}

Similarly we may deal with the case
\[
f^i(R^i)=-\frac{R^i}{n},\quad n=1,2,\ldots,
\]
which leads to
\[
G(\lambda,\bR)=\prod_{i=1}^N\Big(\lambda+\frac{R^i}{n}\Big)^{n}.
\]
Now to generate the general solution  of \eqref{3.24} we take
for
each $i=1,\ldots,N$  a path $\gamma_i(R^i)$ in the complex $\lambda$-plane
ending at  $\lambda_i=-R^i/n$ so that we can write
\begin{equation*}
\Phi:=\frac{1}{2\pi i}\sum_{i=1}^N\int_{\gamma_i(R^i)}
\phi_i(\lambda) G(\lambda,\bR)\d \lambda,
\end{equation*}
where the functions $\phi_i(\lambda)$ are arbitrary.

Let us
consider in detail the case
\[
f^i(R^i)=-R^i,\quad i=1,2.
\]
One finds
\[
\Phi=\Big(\theta_1(-R^1)-\theta_2(-R^2)\Big)(R^1-R^2)+
2\int^{-R^1}\theta_1(\lambda)\d \lambda+2\int^{-R^2}\theta_2(\lambda)\d \lambda,
\]
with $\theta_i(\lambda)$ being arbitrary functions. A normalized
generating function is given by
\begin{equation}\label{3.26a}
G(\lambda,\bR)=\frac{(\lambda+R^1)(\lambda+R^2)}{(\lambda+\lambda_0)^2},
\quad \lambda_0\neq 0.
\end{equation}

We may characterize the general solution of the hydrodynamic flows
\eqref{3.19} corresponding to $n=-1$ and $n=-2$ by means of the hodograph
formula
\begin{equation}\label{3.26}
x+A_{-1}(\lambda=-R^i)\,y+A_{-2}(\lambda=-R^i)\,z=D_i\Phi,\quad i=1,2,
\end{equation}
where $y:=x_{-1},\; z:=x_{-2}$. From \eqref{3.26a} one calculates
\[
b_0=c^2R^1R^2,\;\;\; b_1=c^2(R^1+R^2)
-2c^3R^1R^2,\quad
c:=\frac{1}{\lambda_0},
\]
and gets that the system \eqref{3.26} reads
\begin{align}
\nonumber
(x-c^2z)-c^2(y-2cz)R^2=\dot{\theta}_1(-R^1)(R^2-R^1)-\theta_1(-R^1)
-\theta_2(-R^2),\\
\label{3.27}
\\\nonumber
(x-c^2z)-c^2(y-2cz)R^1=\dot{\theta}_2(-R^2)(R^1-R^2)-\theta_1(-R^1)
-\theta_2(-R^2).
\end{align}
In particular, it implies
\[
\bR=\bR(x-c^2z,y-2cz),
\]
so that $\bR$ is constant on the  straight lines
\[
\vec{x}:=(x,y,z)=(x_0,y_0,0)+(c^2+2c,1)s.
\]
We notice that starting from these solutions we may generate
solutions
\[
u(x,y,z):=\int^y b_0(\bR)\d y+\int^z b_1(\bR) \d z,
\]
of the nonlinear equation
\[
u_{yy}=u_uu_{zx}-u_{yx}u_z.
\]

\vspace{0.3cm}
\noindent {\bf Acknowledgements}
\vspace{0.3cm}

The authors thank Professor Eugeni Ferapontov for useful discussions.
A.B. Shabat was supported by a grant of the Ministerio de Cultura
y Deporte of Spain. L. Martinez Alonso was supported by the DGCYT
project BFM2002-01607.

\end{document}